\documentclass[letters,a4paper,fleqn,usenatbib]{mnras}

\usepackage{mathptmx}

\usepackage[T1]{fontenc}
\usepackage{ae,aecompl}

\usepackage{graphicx}	
\usepackage{amsmath}	
\usepackage{amssymb}	

\newcommand {\hi} {{\rm H}\,{\small\rm I}}
\newcommand {\hicap} {{\rm H}\,{\scriptsize\rm I}}
\newcommand {\hii} {{\rm H}\,{\small\rm II}}

\newcommand {\kms} {\,{\rm km\,s}^{-1}}

\newcommand {\erg} {\,{\rm erg}}
\newcommand {\pc} {\,{\rm pc}}

\newcommand {\kpc} {\,{\rm kpc}}

\newcommand {\cmmc}{\,{\rm cm^{-3}}}

\newcommand {\de}{^{\circ}}
\newcommand{\rsun}{\,{\rm R}_\odot}

\newcommand {\msun}{\,{\rm M}_\odot}

\newcommand {\msunsqkpcyr}{\,{\rm M}_\odot \, {\rm kpc}^{-2} \, {\rm yr}^{-1}}
\newcommand{\vsun}{\,{v}_\odot}
\newcommand{\zsun}{\,{Z}_\odot}

\newcommand{\Myr}{\,{\rm Myr}}

\newcommand{\K}{\,{\rm K}}

\newcommand {\vlsr}{v_{\rm LSR}}

\title[On the origin of the Smith Cloud]{The Galactic fountain as an origin for the Smith Cloud}

\author[A. Marasco and F. Fraternali]{
A. Marasco$^{1}$\thanks{E-mail: marasco@astro.rug.nl}
and F. Fraternali$^{1,2}$
\\
$^{1}$Kapteyn Astronomical Institute,  Postbus 800, 9700 AV, Groningen, The Netherlands\\
$^{2}$Department of Physics \& Astronomy, University of Bologna, via Berti Pichat 6/b, 40127, Bologna, Italy
}

\date{Accepted XXX. Received YYY; in original form ZZZ}

\pubyear{2016}

\begin{document}
\label{firstpage}
\pagerange{\pageref{firstpage}--\pageref{lastpage}}
\maketitle

\begin{abstract}
The recent discovery of an enriched metallicity for the Smith high-velocity \hicap\ cloud (SC) lends support to a Galactic origin for this system.
We use a dynamical model of the galactic fountain to reproduce the observed properties of the SC.
In our model, fountain clouds are ejected from the region of the disc spiral arms and move through the halo interacting with a pre-existing hot corona.
We find that a simple model where cold gas outflows vertically from the Perseus spiral arm reproduces the kinematics and the distance of the SC, but is in disagreement with the cloud's cometary morphology, if this is produced by ram-pressure stripping by the ambient gas.
To explain the cloud morphology we explore two scenarios: a) the outflow is inclined with respect to the vertical direction; b) the cloud is entrained by a fast wind that escapes an underlying superbubble.
Solutions in agreement with all observational constraints can be found for both cases, the former requires outflow angles $\!>\!40\de$ while the latter requires $\gtrsim1000\kms$ winds.
All scenarios predict that the SC is in the ascending phase of its trajectory and have large - but not implausible - energy requirements.

\end{abstract}

\begin{keywords}
ISM: bubbles -- ISM: clouds -- ISM: jets and outflows -- Galaxy: halo 
\end{keywords}



\section{Introduction}
High-velocity clouds \citep[HVCs,][]{WakkervanWoerden97} are large complexes of multiphase gas whose position-velocity is incompatible with them being part of the Galaxy disc.
Their origin has been debated since the moment of their discovery, with two alternative scenarios proposed.
One possibility is that HVCs have an extragalactic origin, either as gas stripped from satellites \citep{Putman+03} or as pristine material inflowing from the intergalactic space \citep{Blitz+99}.
In this scenario the HVCs are currently accreting onto the Galaxy, building-up the gas reservoir that is consumed by the process of star formation. 
The alternative is a Galactic origin, where HVCs participate to a galactic-scale gas cycle triggered by stellar feedback, the so-called galactic fountain \citep{Bregman80, Fraternali+15}.
An accurate determination of distances and metallicities of the HVCs is the key to disentangle between the two scenarios.

The Smith Cloud \citep[SC,][]{Smith63} is one of the best-studied HVCs.
It is located around $l,b\simeq39\de,-13\de$ at $\vlsr$ of about $+100\kms$, has total \hi\ mass of $\sim10^6\msun$ distributed in a coherent structure of $1\times3\kpc^2$ \citep{Lockman+08}, and a similar \hii\ mass \citep{Hill+09}.
Its distance from the Sun ($9.8-15.1\kpc$) has been determined by \citet{Wakker+08} via absorption line studies of background and foreground sources. 
The SC has a head-tail morphology, with the head closer to the midplane, which suggests an ongoing interaction with the ambient medium.
Different origins have been proposed for this system, such as a magnetised \hi\ jet from the 4-\kpc\ molecular ring of the disc \citep{Sofue+04}, or as a gaseous remnant of a dwarf galaxy like the Sagittarius dwarf \citep{BlandHawthorn98}.

Recently, \citet{Fox+16} estimated the metal abundance of the SC by studying the absorption line spectra from three active galactic nuclei lying in the background of the system.
They found a mean metallicity of $0.53$ Solar, which strongly supports a Galactic origin for the SC.
Of the three absorption features, only one overlaps clearly with the \hi\ emission of the SC and shows a metallicity of $\sim0.7$ Solar, while the others ($Z\sim0.8,0.3$ Solar) are quite distant and potentially not associated with the Cloud. 
For this reason, we speculate that the SC is more metal enriched that what determined by \citet{Fox+16}.

\citet[][hereafter F15]{Fraternali+15} proposed a model of the galactic fountain to explain the properties of another well-known HVC, complex C.
In their model, complex C has formed by a powerful gas ejection from the disc in the region of Cygnus spiral arm.
The ejection triggered the condensation of a vast portion of metal-poor coronal gas that, mixing with the enriched material from the disc, lowered the metal abundance of the complex down to the observed sub-solar value \citep[$0.1\!-\!0.3\zsun$,][]{Collins+07}.
In this Letter we show that this model is also applicable to the SC. 

\section{Methods}
As in F15, we use a dynamical model of the galactic fountain to follow the trajectory of fountain clouds through the Galactic halo.
In this model, the cloud motion is governed by three forces.
The first is gravity, that we model by using the Galactic potential of \citet{gd08} (Model I).
The others result from the interaction of the cloud with the Galactic hot corona, and are the drag force and another force produced by the corona condensing onto the cloud's turbulent wake.
Coronal gas has a constant density of $10^{-3}\cmmc$ and rotates around the Galactic centre with a lag of $75\kms$ with respect to the local circular velocity \citep{Marinacci+11}.
The condensation is modelled as an inelastic collision process, where the growth of the cloud mass with time is described by the hydrodynamical simulation presented by F15 and governed by the parameter $t_{\rm delay}$, which we fix to the best-fit value found by F15 for complex C ($46\Myr$).
The deceleration due to drag is given by $\pi R_{\rm cl}^2 \rho_{\rm h} M_{\rm cl}^{-1} v_{\rm rel}^2$, being $M_{\rm cl}$ and $R_{\rm cl}$ the cloud mass and radius, $\rho_{\rm h}$ the coronal gas density and $v_{\rm rel}$ the cloud-corona relative-velocity.
$v_{\rm rel}$ is computed run-time during the cloud's orbit, while the other quantities are fixed to the values used in the simulation ($M_{\rm cl}\!=\!10^5\msun$, $R_{\rm cl}\!=\!180\pc$\footnote{As in F15, we assume that the SC is made up by tens of such clouds.}). 
We discuss the impact of different choices for the drag and condensation parameters in Section \ref{conclusions}. 

We assume that fountain clouds are ejected from the regions of the Galaxy's spiral arms, where star formation is more prominent, and consider the four-arm spiral pattern model of \citet[][see also Fig.\,\ref{orbits}]{Steiman-Cameron+10} and an arm pattern speed of $25\kms\,{\rm kpc}^{-1}$ \citep{Gerhard11}.
The Sun is approximately at the co-rotation radius, given the Galactic constants of $\rsun\!=\!8.5\kpc$ and $\vsun\!=\!220\kms$.

The main observational constrain of our model is the LAB 21-cm all-sky survey of the Milky Way \citep{Kalberla+05}, from which we extract a region containing the \hi\ emission from the SC ($30\de\!<\!l\!<\!60\de$, $-41\de\!<\!b\!<\!-4\de$ and $76\!<\!v_{\rm LSR}\!<\!144\kms$).
Channels at $v_{\rm LSR}\!<\!76\kms$ are discarded as the emission from the SC blends with that of the Galactic disc.
A second observational constraint is the distance $d_{\rm S}$ of the SC from the Sun as derived by \citet{Wakker+08}, $9.8\!<\!d_{\rm S}\!<\!15.1\kpc$ or, considering this range as a $2\sigma$ confidence measurement, $d_{\rm S}\!=\!12.45\pm1.32\kpc$.

Our modelling is performed in two steps.
The first is the `orbit-fitting' routine, a brute-force exploration of the parameter space to find families of orbits with properties compatible with those of the SC.
This allows us to check for the presence of multiple solutions and to study the degeneracy of the parameters.
We characterise the SC by 8 quantities: its average Galactic coordinates ($l_{\rm S}\!=\!41.1\de$ and $b_{\rm S}\!=\!-16.2\de$), its line-of-sight velocity $v_{\rm S}\!=\!96.5\kms$, its distance from the Sun $d_{\rm S}$ and the errors associated to these values ($\delta l_{\rm S}\!=\!3.3\de$, $\delta b_{\rm S}\!=\!3.4\de$, $\delta v_{\rm S}\!=\!12.9\kms$ and $\delta d_{\rm S}$).
Aside from the distance, these values are measured directly on the LAB datacube as \hi-weighted mean and standard deviation quantities.
For each spiral arm, we follow the trajectory of $10^6$ clouds ejected from the arm with random initial conditions, which are the free parameters of our model and vary depending on the details of the model considered (see Section \ref{results}).
At each timestep ($\sim0.3\Myr$) we project the 3D position and velocity of the cloud to a rotating Galactic coordinate systems, which gives the $l$, $b$, $v_{\rm LSR}$ and $d$ of the cloud at that time, and compute the chi-square
\begin{equation}\label{chisquare}
\chi^2 = \left(\frac{l-l_{\rm S}}{\delta l_{\rm S}}\right)^2 + \left(\frac{b-b_{\rm S}}{\delta b_{\rm S}}\right)^2 + \left(\frac{v_{\rm LSR}-v_{\rm S}}{\delta v_{\rm S}}\right)^2 + \left(\frac{d-d_{\rm S}}{\delta d_{\rm S}}\right)^2.
\end{equation}
For every orbit we record the coordinates and the time $t_k$ that give the lowest $\chi^2$.

The second step of our modelling is the `datacube-fitting' routine, where we refine the parameters of our best-fit orbits by fitting synthetic \hi\ observations to the LAB data.
This ensures that our models are consistent with the whole \hi\ position-velocity distribution of the SC.
We focus on the orbit with the lowest $\chi^2$ and use its parameters as initial guess for our fit.
Here, the galactic fountain is modelled as a collection of clouds that have been ejected from a limited region of a spiral arm at a look-back time $t_k$ and for a time duration $\Delta t_k$.
Such a model has the same free parameters as in the orbit-fitting routine, plus $t_k$, $\Delta t_k$, and the extent of the ejection region.
A synthetic \hi\ datacube is produced for any given choice of this parameter set, and the model is fit to the LAB data by minimising the residual |model-data| via the downhill simplex method \citep{Press+02}.
F15 provide additional details on this procedure.

\section{Results} \label{results}
\subsection{Base models with vertical outflows} \label{base}
We first consider a base model where fountain clouds are ejected perpendicularly to the Galaxy midplane.
This model has only two free parameters: the galacto-centric radius of the ejection $R_k$, free to vary between $0$ and $13.5\kpc$, and the kick velocity $v_k$, which varies between $100$ and $400\kms$.
Table \ref{orbitsfit1} lists the optimal parameters and the minimum $\chi^2$ derived for this model for each spiral arm separately.
The quoted $1\sigma$ errors-bar on the parameters accounts for orbits with $(\chi^2\!-\!\chi^2_{\rm min})\!<\!1$ \citep[see][]{Press+02}.
The overall best-fit solution ($\chi^2\!=\!0.03$) comes from the Perseus arm, and it consists of an high-speed outflow ($v_k\!=\!332\kms$) that has occurred $11\Myr$ ago around $R_k\!=\!7.7\kpc$ ($l\!=\!40\de$).
The solutions for the other arms have a much higher $\chi^2$.
Note that the optimal $t_k$ is in the range $10-25\Myr$ for all spiral arms, thus the best agreement with the observed kinematics is always achieved during the ascending part of the cloud's orbit.
Hence, our results indicate that the SC is escaping the disc rather than infalling onto it.

We focus on the Perseus Arm and refine our model parameters by fitting synthetic observations to the LAB datacube.
In the top row of Fig.\,\ref{channelmaps} we compare five representative channel maps extracted from the LAB data around the location of the SC (highlighted with blue contours) with those predicted by our best-fit synthetic observation (orange contours).
The overall agreement with the data is good, especially at $v_{\rm LSR}\!\ge\!107\kms$.
The projected trajectory of the cloud for this model is shown by the orange line in Fig.\,\ref{orbits}.
Clearly, the cloud is still in the early ascending phase of its orbit.
The cloud's future trajectory is very uncertain, given that the condensation of coronal gas can be significant at later times. 
An accurate parametrisation of this process would require dedicated hydrodynamical simulations like those in F15 but it is beyond the scope of this Letter.

\begin{table}
	\centering
	\begin{minipage}{84mm}
		\caption{Best-fit parameters for a base model with vertical outflows as derived from our orbit-fitting routine.}
		\label{orbitsfit1}
		\begin{tabular}{llcccc}
		\hline
			& & Crux & Carina & Perseus & Cygnus \\
		\hline
		$R_k\,^a$		&$\kpc$	&	$4.4\pm0.1$	&	$5.7\pm0.1$	&	$7.7\pm0.1$	&	$9.6\pm0.2$	\\
		$v_k\,^b$  		&$\kms$	&	$198\pm27$	&	$189\pm25$	&	$332\pm58$	&	$>400$	\\
		\hline
		$t_k\,^c$  		&$\Myr$	&	$24\pm1$		&	$20\pm1$		&	$11\pm3$		&	$12\pm1$	\\
		$\chi^2_{\rm min}$	&		&	$13.5$		&	$4.7$		&	$0.03$		&	$4.1$	\\
		\hline
		\end{tabular}\\
	$^a$galacto-centric kick radius; $^b$kick velocity; $^c$orbital time.\\
	\end{minipage}
\end{table}

\begin{table}
	\centering
	\begin{minipage}{84mm}
	\caption{As for Table \ref{orbitsfit1}, but for for a model of inclined outflow. The orientation of the cloud-corona relative motion is considered in the estimate of the goodness of the fit.}
	\label{orbitsfit2}
	\begin{tabular}{llcccc}
		\hline
			& & Crux & Carina & Perseus & Cygnus \\
		\hline
		$R_k$ 			&$\kpc$	&	$4.2\pm0.1$	&	$5.8\pm0.2$	&	$8.7\pm0.2$	&	$11.1\pm0.3$\\
		$v_k$  			&$\kms$	&	$191\pm12$	&	$181\pm20$	&	$183\pm15$	&	$208\pm29$\\
		$\theta_k\,^a$  		&$\de$	&	$1\pm1$		&	$17\pm4$		&	$43\pm4.0$	&	$51\pm6$	\\ 
		$\phi_k\,^b$  			&$\de$	&	$249\pm19$	&	$132\pm11$	&	$165\pm8$	&	$175\pm9$\\ 
		\hline
		$t_k$  			&$\Myr$	&	$24\pm1$		&	$31\pm2$		&	$40\pm1$		&	$49\pm1$	\\
		$\chi^2_{*,\rm min}$	&		&	$16.5$		&	$2.1$		&	$0.2$		&	$0.7$	\\
		\hline
	\end{tabular}\\
	$^a$kick inclination angle; $^b$kick direction angle.\\
	\end{minipage}
\end{table}

\subsection{Accounting for the morphology of the SC}
\begin{figure*}
	\includegraphics[width=0.80\textwidth]{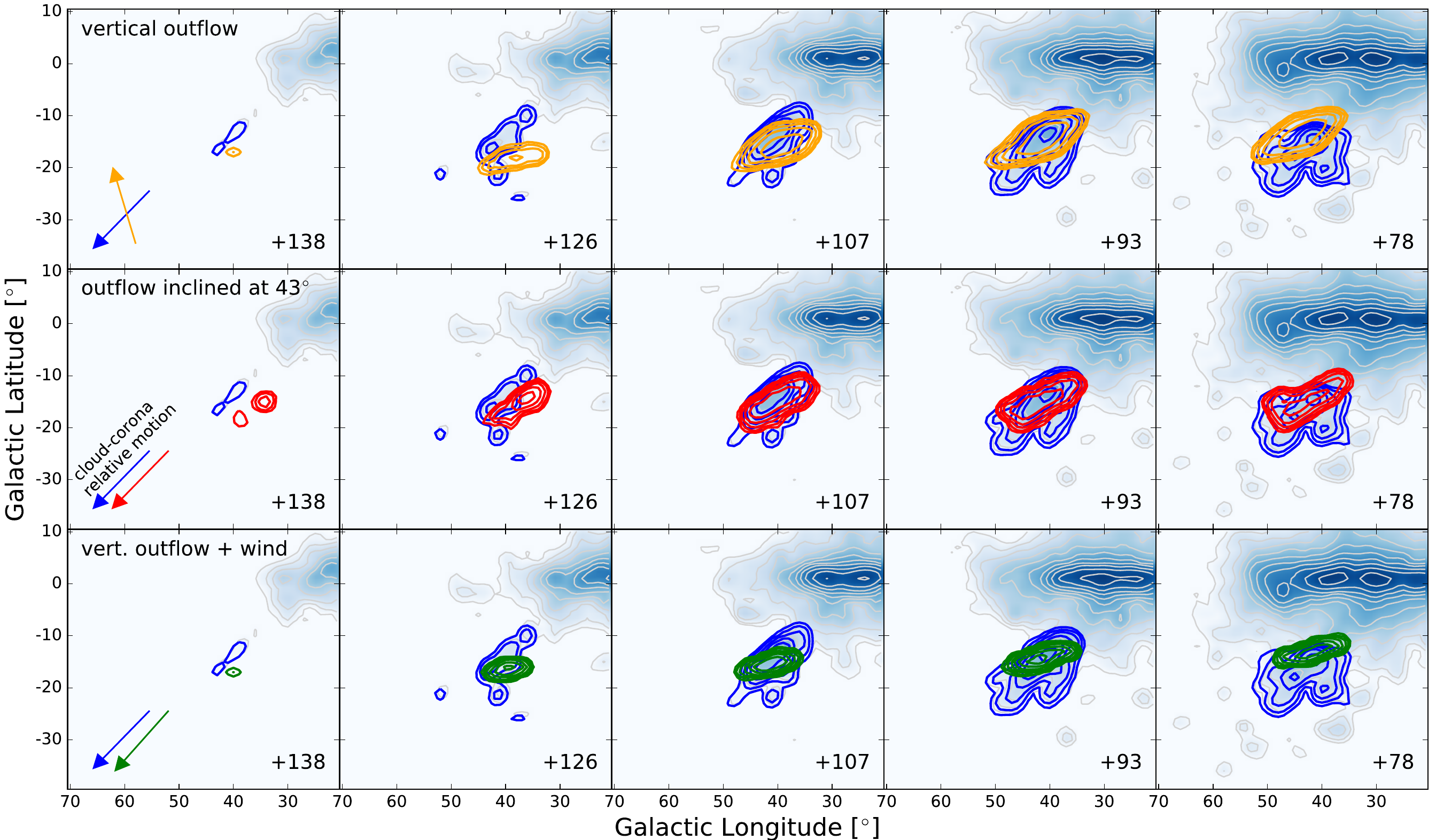}
    \caption{Five representative channel maps from the \hicap\ LAB survey (blue shades, grey contours) in the region of the SC (blue contours) at $v_{\rm LSR}$ of $138$, $126$, $107$, $93$ and $78\kms$ (from the left to the right). \hicap\ emission from the SC at lower $v_{\rm LSR}$ blends with that of the Galaxy disc and is not shown. From top to bottom, the three rows show the \hicap\ emission predicted by different models of the galactic fountain: a vertical outflow from the Perseus Arm (orange contours), an outflow inclined at $43\de$ (red contours) and a vertical outflow in the presence of a fast superbubble wind (green contours). Arrows show the direction of the coronal gas flow in the reference frame of the cloud for both the data (blue arrow, inferred from the SC's morphology) and the models (other colours). Models and data are smoothed to $2\de$ of angular resolution. Brightness temperature contours range from $0.05$ to $51.2\K$ scaling by a factor of 2.}
    \label{channelmaps}
\end{figure*}

The head-tail morphology of the SC suggests an ongoing interaction between this system and the surrounding gas.
Interpreting this morphology as due to ram pressure by the Galactic hot corona gives a constraint on the direction of the SC-corona relative motion projected on the plane of the sky at the current time. 
We use the high-resolution \hi\ map of the SC shown by \citet{Lockman+08} to determine the direction angle $\psi$ by which, in the reference frame of the SC, the coronal gas flows.
We find $\psi\!=\!133\de$, measured clockwise from the longitude axis (blue arrows in Fig.\,\ref{channelmaps}). 
We stress that this is only the \emph{projected} orientation of the 3D SC-corona relative motion, and one should be cautious before concluding that the SC currently moves towards the Galactic disc as both geometrical effects and the presence of a spinning corona complicates the picture.
Our base model of vertical outflow from the Perseus Arm predicts $\psi\sim254\de$ (orange arrow in Fig.\,\ref{channelmaps}), so this model is inconsistent with the morphology of the SC.

We now use $\psi$ as an additional constraint to our model. 
In the orbit-fitting routine we consider a new chi-square $\chi_*^2\!\equiv\!\chi^2+(\psi_{\rm S}-\psi)^2/\delta\psi^2$, where $\chi^2$ is given by eq.\,(\ref{chisquare}), $\psi_{\rm S}\!=\!133\de$ and we assume an ad-hoc $\delta\psi\!=\!15\de$.
We find that, for our base model, the orbits that minimize the new estimators have all $\chi_*^2\!>\!15$. 
Thus a model in which fountain clouds are ejected vertically from the spiral arms cannot explain the kinematics \emph{and} the morphology of the SC simultaneously.

\begin{figure}
	\includegraphics[width=0.50\textwidth]{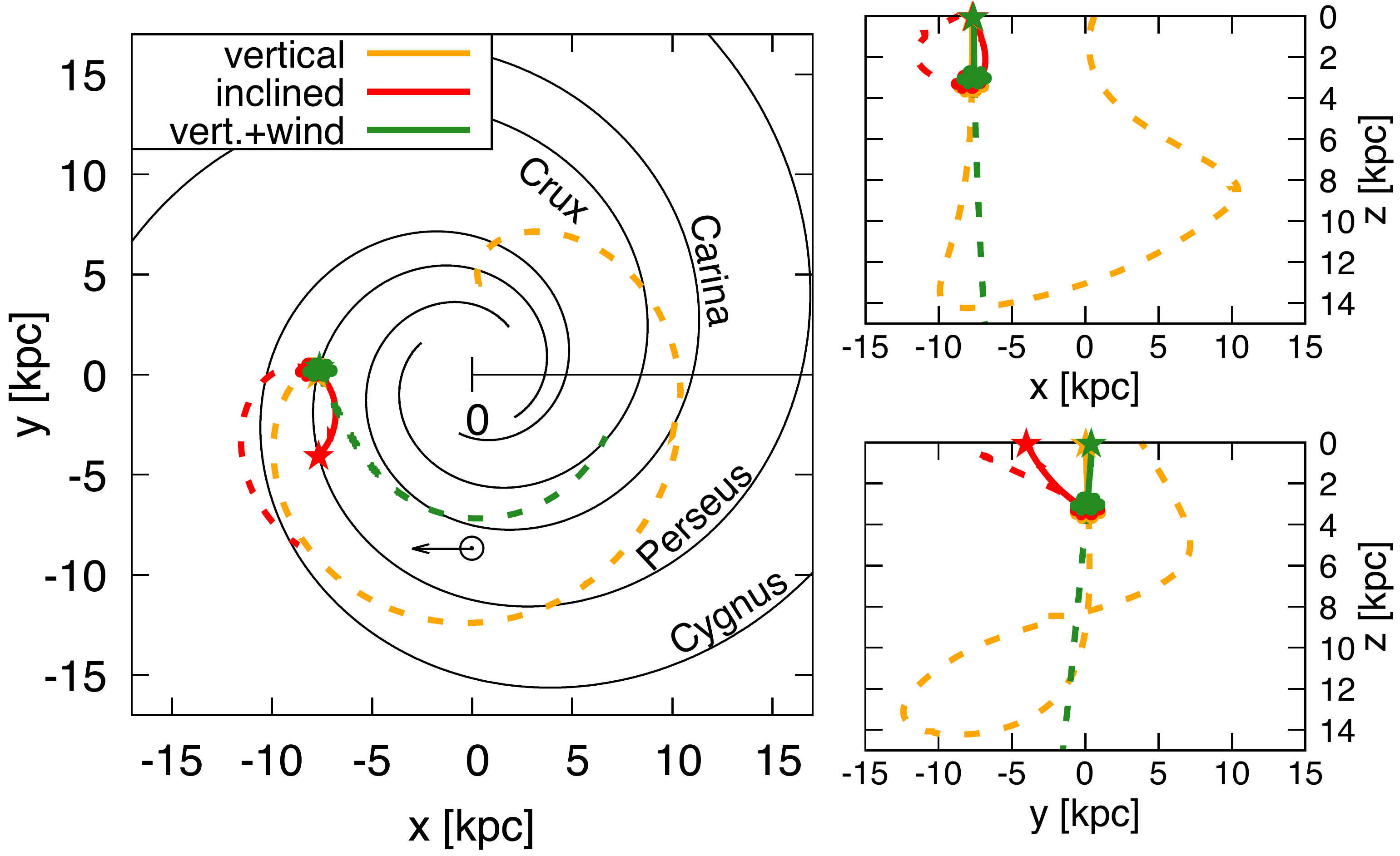}
    \caption{Face-on (left panel) and edge-on (right panels) projections of the best-fit orbits for the three models of the galactic fountain shown in Fig.\,\ref{channelmaps}. Orbits are shown as seen from a reference frame that co-rotate with the Sun ($\odot$). Solid (dashed) lines show the past (future) trajectory of the cloud. Star and cloud symbols indicate the region of the outflow and the current position of the SC, respectively. The face-on view shows the four-arm spiral pattern of \citet{Steiman-Cameron+10}.}
    \label{orbits}
\end{figure}

Although superbubbles should expand preferentially perpendicularly to the disc, where the pressure gradient is maximum, departures from the vertical direction are plausible as the shape of a superbubble is stochastic.
We refine our model by relaxing the hypothesis of vertical kick and allowing the fountain clouds to be ejected in a randomly chosen direction.
The new model has four free parameters: $R_k$, $v_k$, the kick inclination angle $0\!<\!\theta_k\!<\!70\de$ (assumed to be 0 for vertical kicks), and the kick direction angle $0\!<\!\phi_k\!<\!360\de$ (measured clockwise from below in the face-on map of Fig.\,\ref{orbits}).

Table \ref{orbitsfit2} lists the best-fit parameters for this model. 
Different arms can now provide distinct solutions with a low $\chi_*^2$.
The solution found for the Cygnus arm consists of a highly inclined outflow occurring at the periphery of the star-forming disc, which may be unlikely, while in the best-fit orbit found for the Carina arm, $\psi$ is consistent with the observations only for a very narrow time window ($\sim0.3\Myr$).
Hence, we favour the solution found for the Perseus arm: an outflow with $v_k\!=\!183\kms$ and $\theta_k\!=\!43\de$ that has occurred $40\Myr$ ago around $R_k\!=\!8.7\kpc$ ($l\!=\!60\de$).
The ($x,z$) edge-on projection of the cloud's orbit (red line in the top-right panel of Fig.\,\ref{orbits}) reveals that the cloud is approaching the turning point of its trajectory, but it is still in the ascending phase.
However, because of projection effects, in the $(l,b)$ plane the cloud has already passed the turning point and is currently moving towards lower latitudes.
The best-fit synthetic observation is shown in the middle row of Fig.\,\ref{channelmaps} (red contours).
This model performs slightly better than the previous one in the highest velocity channels and predicts a direction for the coronal gas flow that is in excellent agreement with the data. 
The gas that is pushed away from the main body of the SC by ram pressure would have a $v_{\rm LSR}$ lower than that of Cloud, in line with the observed head-tail velocity gradient \citep{Lockman+08}.

\subsection{The effect of a superbubble wind}
We now consider a scenario where the outflowing SC is entrained by a fast wind that escapes the underlying superbubble: ram pressure stripping by the wind would produce the observed cometary shape.
This scenario is corroborated by high resolution hydrodynamical simulations of the ISM, where cometary-like features of cold gas entrained by a hot wind can be seen above newly-born star clusters \citep[e.g. Fig.\,2 in][]{Gatto+16}.

In general, hydrodynamical simulations are needed to model in detail the interactions between the wind, the ISM and the circumgalactic medium.
Here, we include in our dynamical model a simple parametric prescription for a wind that expands uniformly from the Galaxy disc.
The wind is modelled as a gas layer that interacts and mixes with the pre-existing corona, altering the density and the kinematics of the latter.
We consider a cylindrical geometry where the wind moves perpendicularly to the disc at speed $v_w$ and has a Gaussian density distribution in $z$, with midplane density $n_w$ and scale height $\sigma_w$.
Galaxy superwinds have typical densities of $\sim0.1\cmmc$ and velocities up to $2500\kms$ \citep{StricklandHeckman09}. 
We regard these values as upper limits for our model and confine the parameter space to the range $300<v_w<2500\kms$, $-3<\log(n_w)$<-1.0, and assume $0.1<\sigma_w<5\kpc$. 
The kinematics of the system corona+wind are derived by assuming that the two gas layers exchange momentum via inelastic collisions, while the density is given by the sum of the two components.
We consider a constant $v_k$ of $80\kms$ for the cloud, representative for the expansion of the superbubble shell, and leave to the wind the duty of accelerating the cloud via drag.

We include our wind prescription into our model of vertical outflow and search for the best-fit orbits in the Perseus Arm around $R_k\!=\!7.7\kpc$. 
We find that, while the wind scale height is well constrained ($\sigma_w\!=1.5\pm0.6\kpc\!$), $n_w$ and $v_w$ are degenerate and highly anti-correlated.
We fix $n_w$ to $0.01\cmmc$ and find the best-fit orbit ($\chi_*^2\!=\!0.13$) at $v_w\!=\!1200\pm139\kms$ for very short orbit times $t_k\!=\!6\pm2\Myr$.
Assuming $n_w\!=\!0.1\cmmc$ gives $v_w\sim950\kms$, thus in all cases high-speed winds are required to reproduce the properties of the SC.
The resulting best synthetic observation for this model is shown in the third row of Fig.\,\ref{channelmaps}.
The agreement with the data is similar to that shown by our base model, but now the relative motion of the ambient gas points toward the right direction (green arrow in Fig.\,\ref{channelmaps}).
The cloud's trajectory (green lines in Fig.\,\ref{orbits}) in the first few million years is similar to that of our base model, but at later times the wind pushes the cloud at much larger heights above the disc (not shown in Fig.\,\ref{orbits}). 
Surely the future of this cloud cannot be predicted by our simple model.

\section{Discussion and conclusions}\label{conclusions}
Inspired by the findings of \citet{Fox+16}, in this Letter we have investigated the possibility that the Smith high-velocity \hi\ Cloud has originated via a Galactic fountain from the regions of the spiral arms of the Milky Way.
We used a dynamical model of the galactic fountain and focussed on those orbits that reproduce the main observational properties of the SC, such as its position, line-of-sight velocity, distance and head-tail morphology.
We found that a simple model of vertical outflow from the Perseus Arm, from a Galactocentric distance of $\sim7.7\kpc$ at a speed of $\sim330\kms$, reproduces all these constraints except the morphology, if we interpret it as due to ram-pressure stripping by the corona.
The same model can be refined by including a fast ($\gtrsim1000\kms$) wind that escapes the underlying superbubble and entrains the cloud, providing the ram pressure required. 
Alternatively, a model where the outflow is inclined by $\sim43\de$ can reproduce all the observational constraints.

In all scenarios considered, the SC is in the ascending phase of its trajectory and has travelled for no more than $50\Myr$.
Such a short orbital time implies that the coronal gas condensation is minimal - less than $4\%$ of the final cloud mass comes from the corona - thus the cloud maintains the same metallicity of the underlying disc, in line with the findings of \citet{Fox+16}.
We have verified that, in all scenarios, increasing the corona condensation rate (i.e. $t_{\rm delay}$) systematically worsens the fit.

In the models without wind the typical drag timescale is $\sim\!150\Myr$.
This is much longer than $t_k$, thus drag has little impact on the cloud trajectories.
Decreasing this timescale to values closer to $t_k$ drastically worsens the fit. 
In the wind model the drag plays a key role in accelerating the cloud, thus varying the drag parameters has some influence: for $M_{\rm cl}\!=\!10^4\msun$ ($10^6\msun$) the best-fit orbits have $v_w\!\sim\!1000\kms$ ($2000\kms$) and $h_s\!\sim\!2.2\kpc$ ($1.3\kpc$).

A galactic fountain origin would exclude the presence of dark matter associated to the SC. 
The metallicity of the SC is incompatible with its being a `dark dwarf galaxy' (which would have a much lower metal content), unless the minihalo has accreted gas from the Galaxy's ISM during a previous passage through the disc as in the model of \citet{GalyardtShelton16}.

\citet{Lockman+08} computed a trajectory for the SC and found that the system has crossed the Galactic plane from above to below $\sim\!70\Myr$ ago at $R\!=\!13\kpc$.
It is not surprising that we do not recover this trajectory amongst our best-fit solutions, as a) we impose an origin from one of the spiral arms; b) unlike Lockman et al., we do not force the SC's motion to be along the system's major axis.

One could wonder whether the models presented in this work are energetically plausible.
While the \hi\ mass of the SC is well constrained \citep[$\sim10^6\msun$,][]{Lockman+08}, its ionised gas mass is more uncertain and may dominate the total mass budget \citep{Hill+09}.
Assuming a total gas mass in the range $1-5\times10^6\msun$, the kinetic energy $E_k$ required to kick this material at a velocity of $200\kms$ is $0.4\!-\!2\times10^{54}\erg$, comparable to the estimate for Complex C (F15) and to the energy associated to the \hi\ holes observed in the ISM of nearby spirals \citep{Boomsma+08}.
A lower limit for the kinetic energy associated to the wind $E_w$ can be estimated by assuming that the wind operates only in the region of the SC, i.e., it is confined to a cone or a cylinder with base equal to the SC size, for which we use $760\pc$ from the extent of its minor axis.
Adopting $v_w\!=\!1200\kms$ and a Gaussian density distribution in $z$ with $n_w\!=\!0.01\cmmc$ and $\sigma_w\!=\!1.5\kpc$, we find $E_w\!=\!1\!-\!3\times10^{54}\erg$ (depending on the geometry), thus $E_w$ is compatible to $E_k$ found for the model without wind.

The issue is that the time window by which this energy should be released is extremely narrow, as our datacube-fitting routine typically returns $\Delta t_k\!=\!5\Myr$. 
F15 found $\Delta t_k\!=\!50\Myr$ for complex C, and concluded that a star formation rate density (SFRD) of $\sim0.01\msunsqkpcyr$ is needed to lift the complex from the disc to its current location. 
For the SC, the SFRD would be of the order of $0.1\msunsqkpcyr$. This is sufficient to trigger a Galactic wind \citep{Heckman02}.
SFRDs of this magnitude are occasionally measured in nearby galaxies on kpc scales, but they are typically - although not exclusively - associated to the region of the galaxy centre \citep{Leroy+08}.

We conclude that the energy requirements for our galactic fountain models of the SC are improbable, but not impossible.
It is certainly possible that a combination of a superbubble wind and a skewed outflow can provide a good fit to the data and lower the energy requirements at the same time. 
Unfortunately this scenario has too many free parameters to be addressed here.

\section*{Acknowledgements}
AM acknowledges support from the European Research Council under the European Union's Seventh Framework Programme (FP/2007-2013) / ERC Grant Agreement nr. 291531.




\bibliographystyle{mnras}
\bibliography{smithcloud} 





\bsp	
\label{lastpage}
\end{document}